\documentclass[apj]{emulateapj}

\usepackage{color}

\def\gtorder{\mathrel{\raise.3ex\hbox{$>$}\mkern-14mu
             \lower0.6ex\hbox{$\sim$}}}
\def\ltorder{\mathrel{\raise.3ex\hbox{$<$}\mkern-14mu
             \lower0.6ex\hbox{$\sim$}}}

\usepackage{color}
\usepackage{enumitem}
\usepackage{graphicx}	
\usepackage{amsmath}	
\usepackage{bm}		
\usepackage{url}
\usepackage{amsbsy}
\usepackage{hyperref}
\usepackage{longtable}

\usepackage[T1]{fontenc}
\usepackage{ae,aecompl}

\usepackage{amssymb}	
\usepackage{colordvi}
\usepackage{amsmath}
\usepackage{hyperref}
\usepackage{natbib}

\usepackage{academicons}

\usepackage{lineno}

\slugcomment{Submitted to AJ}

\shorttitle{The Oort Cloud}

\shortauthors{Ofek et al.}

\begin{document}

\title{An efficient observational strategy for the detection of the Oort cloud}

\author{
Eran~O.~Ofek\altaffilmark{1,$\star$},
Sarah A. Spitzer\altaffilmark{1},
Guy Nir\altaffilmark{2}
}
\altaffiltext{1}{Department of particle physics and astrophysics, Weizmann Institute of Science, 76100 Rehovot, Israel.}
\altaffiltext{2}{Lawrence Berkeley National Laboratory, 1 Cyclotron Road, MS 50B-4206, Berkeley, CA 94720, USA}
\altaffiltext{$\star$}{Corresponding Author: eran.ofek@weizmann.ac.il}


\begin{abstract}

The Oort cloud is presumably a pristine relic of the Solar System formation.
Detection of the Oort cloud may provide information regarding the stellar environment in
which the Sun was born and on the planetesimal population during the outer planets' formation phase.
The best suggested approach for detecting Oort cloud objects \emph{in situ},
is by searching for sub-second occultations of distant stars by these objects.
Following Brown \& Webster, we discuss the possibility of detecting Oort cloud objects by observing near the quadrature direction.
Due to the Earth's projected velocity, the occultations are longer near the quadrature direction and are therefore easier to detect, but have lower rate.
We show that, for $\lesssim1$\,m size telescopes, the increased exposure time will result in about one to three
orders of magnitude increase in the number of detectable stars 
that have an angular size smaller than the Fresnel scale 
and are therefore suitable for an occultation search.
We discuss the ability of this method to detect Oort cloud objects using existing survey telescopes, and we estimate the detection rate as a function of the power-law index of the size distribution of the Oort cloud objects and their distance from the Sun. We show that occultations detected using $\approx1$-s integration by $\lesssim1$\,m telescopes at the optimal region near the quadrature points will be marginally dominated by Oort cloud objects rather than Kuiper belt objects.

\end{abstract}



\section{Introduction}
\label{sec:intro}

The Oort cloud (\citealt{Oort1950_TheOortCloud}) formation is presumably the result of the scattering of planetesimals from the primordial Solar System, 
where the most efficient scatterers were likely Uranus- and Neptune-mass-like planets (e.g., \citealt{Duncan+1987_OortCloudFormation}).
So far, there are no observations of objects, \textit{in situ}, in the Oort Cloud, 
and their numbers, distances, and size distributions are uncertain.
The Oort cloud content and distance from the Sun may hold clues regarding the Solar System formation and evolution, 
such as the presence of ice planets and their past locations; 
the stellar density at which the Sun was born; 
and the evolution of planetesimals before their ejection from the planets' orbits
(e.g., \citealt{Dones+2004_OortCloudFormation}, \citealt{Brasser+2006_OortCloudFormation_EmbededStarClustersEnvironment}).

\cite{Bailey1976_OccultationsKBO_idea} suggested 
that it is possible to detect small Trans-Neptunian objects by occultations.
However, due to their tiny cross-section and short duration, 
these occultations are hard to find.
So far, only three serendipitous occultations by Kuiper-Belt Objects have been reported (\citealt{Schlichting+Ofek+2009_HST_KBO_Occultations}, \citealt{Schlichting+Ofek+2012_HST_KBO_Occultations}, \citealt{Arimatsu+2019_AkmSizedKBO_Occultation_GroundBasedTelescope}),
as well as a few upper limits (e.g., \citealt{Bianco+2009_OccultationsKBO_MMT_search, Zhang+2013_TAOS_I_Results7years,Nir+Ofek+2023_WFAST_KBO_Search2020_2021, Nir+Ofek+2023_TNO_Occultations_Pipeline}).
The duration of these occultations is controlled  by the Earth's velocity, 
the Fresnel angular size, and the angular sizes of the occulter and the occulted star.
Since the size distributions of Kuiper belt objects, 
and presumably Oort cloud objects, follow a steep power law, 
their population is dominated by small objects.
In this case, the overwhelming majority of the occultations 
is expected from occulters whose sizes are near the Fresnel size (e.g., \citealt{Ofek+2010_OortOccultation_Kepler}).
In such cases,
the semi-duration of Kuiper-Belt occultations
are of the order of 0.05\,s (i.e., Fresnel size $\approx 1$\,km, divided by the Earth's velocity).
For Oort cloud objects at 3000 (1000)\,AU from the Sun, 
the Fresnel size is ten (five) times larger, 
and hence the semi-duration of occultation is of the order of $\sim0.4$ ($0.2$)\,s.

Another critical factor is that for stars whose angular sizes 
are much larger than the angular size of the occulter and the Fresnel size, 
the occultation will be diluted and harder to detect.
Since the angular size of stars depends on the flux and color temperature, 
fainter and bluer stars are preferred.
For example, a Sun-like star at an apparent magnitude of about 14, 
will have an angular radius that is similar to the Fresnel size for objects at 3000\,AU.
Therefore, the detection of occultations by Oort cloud objects 
not only requires sub-second time resolution and a wide field of view, 
but also large telescopes\footnote{Or by combining multiple small telescopes which may be more cost-effective (e.g., \citealt{Ofek+BenAmi2020_Grasp_SkySurvrys_CostEffectivness}).}.
A way around the short exposure requirement 
is to use instruments with long exposure times, but precise photometry.
This technique was suggested for {\it Kepler}- and {\it TESS}-like data by
\cite{Gaudi2004ApJ_Kepler_KuiperBelt_Occultations}
and
\cite{Ofek+2010_OortOccultation_Kepler}.
An alternative option is to use fast readout devices on large telescopes (\citealt{Harding+2016_CHIMERA_P200}).

Here we show that it is possible to increase the efficiency of small telescopes
(e.g., W-FAST [\citealt{Nir+Ofek+2021_WFAST}], Tomo-e Gozen [\citealt{Sako+2018SPIE_Tomo_e_Gozen_Overview}], TAOS-I [\citealt{Zhang+2013_TAOS_I_Results7years}] and TAOS-II [\citealt{Huang+2021PASP_TAOS_II_Occultations_KBOs_Overview}] like systems)
for Oort cloud object detection by about an order of magnitude, using a simple observing strategy.
The idea is to observe the direction near the quadrature 
(90\,deg from the Sun on the ecliptic), 
where the occultation duration is longer,
and a few seconds of integration time is sufficient.
Although the rate of occultation per suitable star will decrease by an order of magnitude, 
the number of suitable stars will increase by about one to three orders of magnitudes.
\cite{Brown+Webster_1997MNRAS_KBO_QuadratureOccultationsDetection} suggested this technique for the detection of KBOs.
They argue that this method may be useful 
because it can be applied to existing sky surveys.
The analysis presented here suggests that for $\lesssim1$\,m telescopes,
the efficiency of this method for Oort cloud object detection, is considerably higher compared to opposition observations.
Furthermore, the expected detection rate of Oort cloud objects near quadrature will be marginally larger than the predicted rate of KBO occultations near quadrature.

In \S\ref{sec:Basic} we provide some background related to Solar System occultations,
and in \S\ref{sec:ExpTime} we discuss exposure times and readout noise considerations.
In \S\ref{sec:Quadrature} we discuss the quadrature method, and we conclude in \S\ref{sec:Conclusion}.

\section{Basics of stellar occultations by Solar System objects}
\label{sec:Basic}

The size distribution of known KBOs follows a power-law
\begin{equation}
    \frac{dN(R)}{dR} = N_{>1\,{\rm km}} \frac{q-1}{1\,{\rm km}} \Big(\frac{R}{1\,{\rm km}}\Big)^{-q}.
\end{equation}
Here $N_{>1\,{\rm km}}$ is the total number of objects with a radius larger than 1\,km. $N(R)$ is the number distribution of objects of radius $R$, and $q$ defines the power-law. 
For KBOs smaller than about 40\,km, $q\approx3.7$ (\citealt{Schlichting+Ofek+2012_HST_KBO_Occultations}).
We assume that a steep power-law size distribution is also relevant for the Oort Cloud. A steep power-law size distribution implies that small bodies dominate the optical depth for occultations.

Solar System occultations have three relevant angular scales.
These are:
(i) the object angular radius $\theta_{\rm obj}=r_{\rm obj}/d$,
where $r_{\rm obj}$ is the object radius, and $d$ is its distance;
(ii) the occulted star's angular radius
\begin{equation}
    \theta_{*} \approx 465
    \Big(\frac{T}{5700\,{\rm K}}\Big)^{-2}
    10^{-0.2 (m_{\rm V} - 4.83)}\,\mu{\rm arcsec}.
\end{equation}
Here we assume the star has a black-body spectrum with
effective temperature $T$, and $V$-band apparent magnitude $m_{\rm V}$ (extinction is discussed in Appendix~\ref{Ap:ext});
and
(iii) the Fresnel radius
($r_{\rm F}$)
and angular Fresnel radius ($\theta_{\rm F}$)
which are:
\begin{equation}
    r_{\rm F} \approx 10.6 \Big(\frac{\lambda}{5000\,{\text \AA}}\Big)^{1/2}
    \Big(\frac{d}{3000\,{\rm AU}}\Big)^{1/2}\,{\rm km},
    \label{eq:rF}
\end{equation}
\begin{equation}
    \theta_{\rm F} \approx 4.9 \Big(\frac{\lambda}{5000\,{\text \AA}}\Big)^{1/2}
    \Big(\frac{d}{3000\,{\rm AU}}\Big)^{-1/2}\,\mu{\rm arcsec},
    \label{eq:thetaF}
\end{equation}
where $\lambda$ is the wavelength at which we observe.

The duration of these occultations is dominated
by the Earth's velocity\footnote{This is not correct near the points of zero angular speed, because of the object's orbital eccentricity.}, 
the Fresnel size, and the angular size of the occulted star and the occulter. 
Assuming a negligible angular size for the star (see below), 
an occulter size smaller than the Fresnel size, 
and observations taken near opposition,
this translates to an occultation semi-duration of about $0.05$
and $0.4$\,s for the Kuiper belt (40\,AU) and Oort cloud (3000\,AU), respectively.
Such short durations require fast readout cameras.

Yet another important challenge is related to the angular size of stars.
To avoid photometric dilution of the occultation, 
the angular size of the star should be smaller than the Fresnel size.
The Fresnel sizes are $41\,\mu$arcsec and $4.6\,\mu$arcsec
for 40\,AU and 3000\,AU, respectively.
A Solar-like star ($T\approx 5700$\,K)
will have these angular sizes at an apparent magnitude ($m_{\rm V})$ of
about 10.1 and 14.9, respectively (see Figure~\ref{fig:ThetaF_T_mV}).
To get a decent $S/N$ for a 15 mag star with sub-second integrations, 
a large collecting area (diameter $\gtorder1$\,m) is required. This means that using the regular (short-exposures) technique, the detection of the Oort cloud requires large telescopes, or $\gtrsim0.5$\,m telescopes with excellent image quality and detectors with low readout noise.

The optical depth of TNOs is very small -- at any given moment, the probability that a star is occulted by a TNO is 
about $10^{-9}$.
Therefore, a good strategy is to target many stars simultaneously.
The alternative is to observe a few stars for a long time (e.g., \citealt{Schlichting+Ofek+2009_HST_KBO_Occultations}).
The short duration of the occultation and the need to observe
a large number of stars means a huge data rate
that in turn requires special data handling and analysis techniques (see e.g., \citealt{Nir+Ofek+2021_WFAST}).
Furthermore, short integrations are affected by intensity scintillation,
which needs to be accounted for in the statistics (e.g., \citealt{Nir+Ofek+2023_TNO_Occultations_Pipeline}).
The next section will discuss another important consideration: the exposure time and the readout noise-dominated regime.

\section{The effect of Exposure time}
\label{sec:ExpTime}

The common strategy for TNOs occultation observations 
is to observe at a high rate, 
typically with exposure times shorter than $0.1$\,s
(e.g., \citealt{Roques+2006_KBO_Occultations_Search}).
When using CCD or CMOS detectors 
with such short exposures, and small to medium-sized telescopes,
the noise is usually dominated by the camera read-noise.
We define the transition between the background-noise dominated regime
and the read-noise-dominated regime 
when the variance of the background is equal to the read-noise squared.
This transition time [s] takes place at an exposure time roughly given by
\begin{equation}
t_{\rm trans} \approx \frac{R^{2}}{p^{2} A 10^{-0.4(b_{\rm V}-14.76)} }\,{\rm s},
\end{equation}
where $R$ is the read-noise (in electrons), $p$ is the pixel scale (in arcsec\,pix$^{-1}$),
$A$ is the effective collecting area of the telescope (in cm$^{2}$), 
and $b_{\rm V}$ is the sky $V$-band magnitude per square-arcsec.
This time scale does not depend on the seeing.
For $R=2$\,e$^{-}$, $p=1$\,arcsec\,pix$^{-1}$, and $b_{V}=21$\,mag\,arcsec$^{-2}$,
this is about $4$\,s, $0.2$\,s, and $0.006$\,s
for 20, 100, and 500\,cm telescopes (diameter), respectively.
%
In the read-noise dominated regime, the Signal to Noise ratio is $S/N\propto t$,
and hence the limiting flux (at some detection threshold) is $F_{\rm lim}\propto t^{-1}$,
while in the background-noise-dominated regime
the limiting flux is $F_{\rm lim}\propto t^{-1/2}$.
This means that in the read-noise-dominated regime, increasing the exposure time increases the number of faint stars faster compared to the
background-noise-dominated regime.
This consideration may play an important role when designing a TNO occultation survey.

When considering short vs.~long integration times, other effects may be important.
For example, when the exposure time is longer than the occultation duration,
the occultation depth will be diluted (e.g., \citealt{Ofek+2010_OortOccultation_Kepler}).
Typically, ground-based surveys' photometric noise is not limited by Poisson noise,
but by other effects like flat-field errors, intensity scintillation (e.g., \citealt{Young1967AJ_Photometry_Scintialtions_ReigerTheoryConfirmation}), and atmospheric absorption.
Therefore, ground-based observations of diluted occultations are not practical in most cases.

\section{The quadrature observations strategy}
\label{sec:Quadrature}

\cite{Brown+Webster_1997MNRAS_KBO_QuadratureOccultationsDetection} suggested utilizing the fact that 
the duration of a TNO occultation depends on its elongation 
(angular distance from the Sun, along the ecliptic),
where the instantaneous duration approaches infinity near the quadrature points.
For a circular orbit on the ecliptic, 
the on-sky speed ($V_{\perp}$), 
as a function of elongation ($\epsilon$) 
is
\begin{equation}
    V_{\perp} = V_{\oplus}\sin{(\epsilon-90)} + V_{\rm obj} \cos{(\sin^{-1}{[a_{\oplus}/a}]\sin{\epsilon})},
    \label{eqn:v_perp}
\end{equation}
where $a$ is the semi-major axis of the object.
This gives points of zero angular velocity (i.e., $V_{\perp}=0$), at angular distances
of 56.3, 80.9, 88.2, and 89\,deg,
for 3, 40, 1000, and 3000\,AU, respectively.

Therefore, observing near the point of zero angular speed 
will result in occultations with a long duration (e.g., larger than a few seconds).
Defining the angular distance from the point of the circular orbit's zero angular speed by $\phi$,
the approximate duration of the occultation increases by a factor of $1/\sin{\phi}$,
compared to occultations observed near opposition.
However, this also means that the occultation rate is proportional to
about $\sin{\phi}$, compared to the occultation rate near opposition.
This seems to reduce the efficiency of this method considerably.
Although the efficiency of this approach at first glance appears to be an order of magnitude lower,
\cite{Brown+Webster_1997MNRAS_KBO_QuadratureOccultationsDetection}
suggested that longer integration times
result in a larger number of monitored stars and are more
practical for KBO detection from a technological point of view -- 
i.e., longer integration time images are easier to obtain and handle using existing equipment.

Here we claim that for Oort cloud object searches using meter-class or smaller telescopes,
this method may be more efficient than opposition observations.
This is because, assuming we are in the read-noise-dominated regime,
increasing the integration time increases the limiting flux of the system linearly.
Hence, the number of detectable stars increases faster than in the background-dominated regime.
Furthermore, when going to fainter magnitudes the average color of stars becomes redder\footnote{This is correct in the magnitude range we are interested in.}.
Therefore,
the fraction of usable stars 
(i.e., smaller than the Fresnel scale),
increases faster than the increase in the number of stars.

\begin{figure}
\includegraphics[width=8cm]{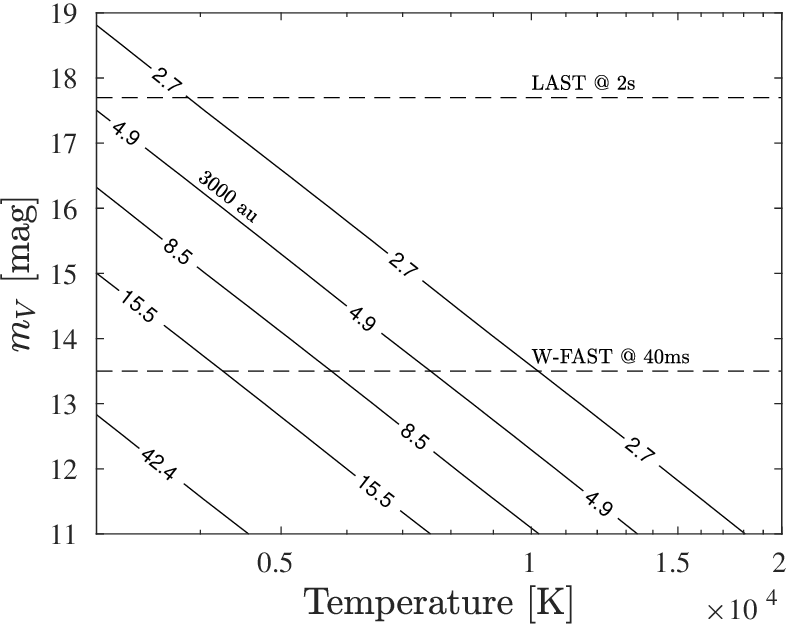}
\caption{The angular radii of stars, in $\mu$arcsec, as a function of their
effective temperature and apparent $V$-band magnitude (ignoring extinction; see Appendix~\ref{Ap:ext}).
The lines correspond to the Fresnel size at 5000\,\AA~and distances of
40, 400, 1000, 3000, $10^{4}$\,AU
(bottom-left to top-right).
The dashed horizontal lines show the approximate limiting magnitude for two surveys:
W-FAST with 40\,ms exposures (\citealt{Nir+Ofek+2021_WFAST}), 
and a single LAST telescope with 2\,s exposure (\citealt{Ofek+2023PASP_LAST_Overview, BenAmi+2023PASP_LAST_Science, Ofek+2023PASP_LAST_PipeplineI}).
\label{fig:ThetaF_T_mV}}
\end{figure}

\begin{figure}
\centering\includegraphics[width=7.5cm]{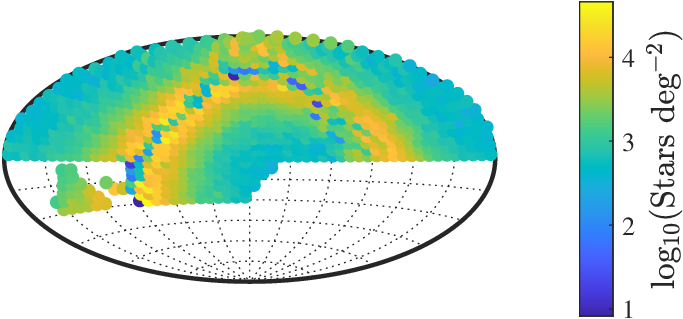}
\caption{The surface density of stars brighter than $g$ magnitude 18 and fitted angular radius smaller than 5\,$\mu$arcsec. Catalog data adopted from Pan-STARRS DR1 catalog (\citealt{Chambers+2016_PS1_Surveys, Schlafly+2012_PS1_PhotometricCalibration}), where the angular sizes were estimated by fitting the $g$ and $r$ magnitudes.
\label{fig:SorcesSmallerThanTF_mag18}}
\end{figure}

\begin{figure}
\includegraphics[width=8cm]{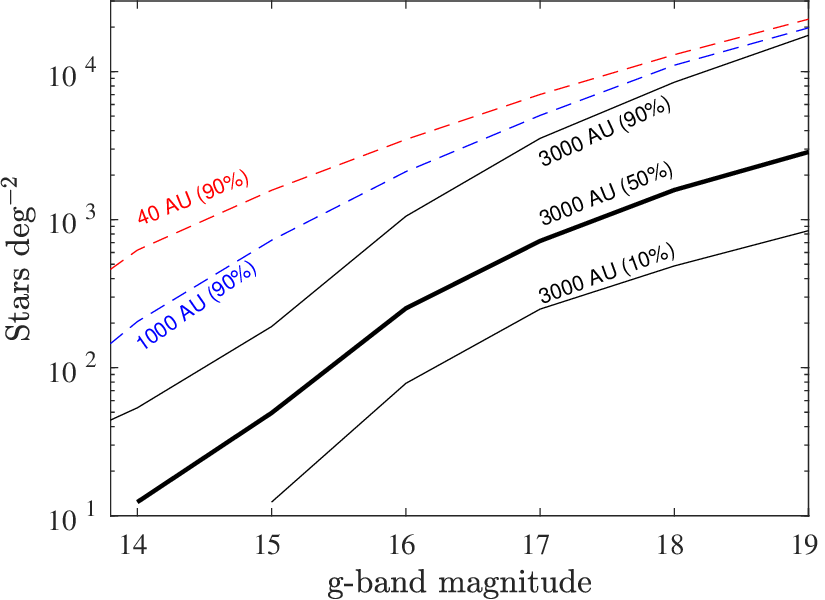}
\caption{The cumulative mean surface density of stars with an angular radius smaller than $5\,\mu$arcsec (i.e., Fresnel radius at 3000\,AU), 
as a function of magnitude (Black solid heavy line).
Thin black lines show the lower 10- and upper 90-percentiles.
The dashed blue (red) lines show the upper 90-percentiles surface density of stars smaller than 8 (42)\,$\mu$arcsec (corresponding to 1000 (40)\,AU).
\label{fig:SorcesSmallerThanTF_vsMag}}
\end{figure}

In Figure~\ref{fig:ThetaF_T_mV}, we show the angular radii of stars ($\mu$arcsec) as a function of their
effective temperature and apparent $V$-band magnitude (ignoring extinction; see Appendix~\ref{Ap:ext}).
To estimate the surface density of stars 
with angular radii smaller than the angular Fresnel radius of Oort cloud objects, 
we used the tools in \cite{Ofek2014_MAAT} and \cite{Soumagnac+Ofek2018_catsHTM}
to fit the black-body temperature and calculate the approximate angular radii
of Pan-STARRS1 sources (\citealt{Chambers+2016_PS1_Surveys}).
Figure~\ref{fig:SorcesSmallerThanTF_mag18} shows 
the surface density of stars brighter than $g$ AB-magnitude 18 
and fitted angular radius smaller than 5\,$\mu$arcsec.
Some sky regions have a stellar density of $10^{4}$\,deg$^{-2}$ brighter than 18 magnitudes
and smaller than about 5\,$\mu$arcsec.
%
Figure~\ref{fig:SorcesSmallerThanTF_vsMag} presents
the surface density of stars with an angular radius smaller than $5$, $8$, and $42$\,$\mu$arcsec, as a function of magnitude.

Next, we would like to estimate the occultation rate of some survey
near the points of zero angular speed.
To simplify the calculations,
here we assume that we are interested
only in stars whose angular size is smaller
than 5\,$\mu$arcsec, and that we cannot detect
an occulter whose radius is smaller than the Fresnel radius.
These limitations allow us to neglect diluted occultations, 
which will be difficult to verify.

In this case, the probability that a given star will be occulted
by an Oort cloud object, per unit time, is roughly given by
\begin{align}
    R_{\rm occ}  \approx \frac{1}{4\pi} 2\theta_{\rm F} \mu N_{>r_{\rm F}} f_{\rm occ} \cr
     \cong 2.2\times10^{-11}\frac{\theta_{\rm F}}{4.9\mu{\rm arcsec}}
    \frac{\mu}{12\mu{\rm arcsec\,s}^{-1}} \frac{N_{>r_{\rm F}}}{10^{11}} f_{\rm occ}\,{\rm s}^{-1}.
    \label{eq:OccRate}
\end{align}
Here $f_{\rm occ}$ is the fraction of occultations
that can be detected using a specific integration time
(i.e., with duration above some minimum time scale),
the angular Fresnel radius $\theta_{\rm F}$ is measured in radians, the number of objects with a radius greater than the Fresnel radius is 
$N_{>r_{\rm F}} = N_{>1\,{\rm km}} (r_{\rm F}/[1\,{\rm km}])^{1-q}$,
and $\mu$ is the angular speed, in radians per second, 
of the objects of interest on the celestial sphere,
approximately given by:
\begin{equation}
    \mu \approx \frac{V_{\oplus}}{a} \sin{\phi} \cong 13.6 \Big(\frac{a}{3000\,{\rm AU}} \Big)^{-1} \sin{\phi} \,\mu{\rm arcsec\,s}^{-1}.    
\end{equation}
However, this expression for $\mu$ is only an approximation.
Specifically, for a more accurate evaluation of $\mu$, 
we need to take into account the object's orbital elements.

To estimate the effect of the non-zero eccentricity and inclinations, 
we present the following simulations.
We generate two billion random orbital elements,
with a constant semi-major axis $a$,
where the longitude of perihelion ($\omega$),
the longitude of ascending node ($\Omega$),
and the mean anomaly, at some fiducial epoch, is distributed uniformly
between 0 and 360\,deg.
The inclination ($I$) is distributed like $\propto \cos{I}$,
and the eccentricity is distributed uniformly between 0 and 0.6.
Next, we calculate the ephemerides of these simulated objects,
at some epoch, for a geocentric observer.
The result of this simulation is a list of Right Ascension and Declination, and on-sky angular speed
for each object.

\begin{figure}
\includegraphics[width=8.5cm]{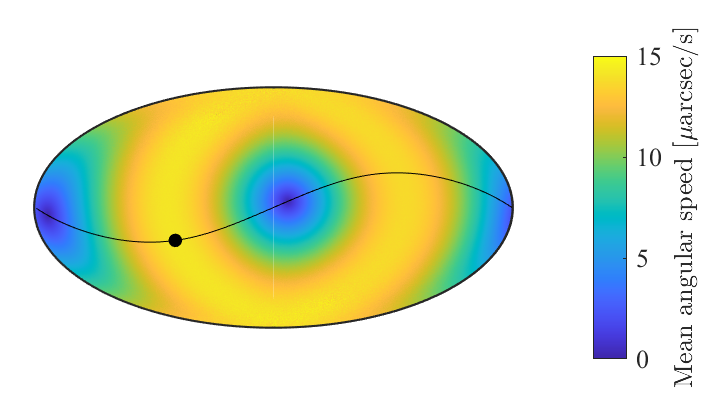}
\caption{Mean angular speed ($\mu$arcsec\,s$^{-1}$) of Oort cloud objects at 3000\,AU 
(with the distribution of the orbital elements specified in the text) as a function of sky position.
The map is shown in equatorial coordinates and Aitoff projection, 
where the origin of the right ascension and declination is at the center. 
The black line shows the ecliptic plane.
The map is plotted at some arbitrary date,
while the black-filled circle shows the Sun position at this time.
The points of minimum angular speed are at 89\,deg from the Sun.
Around the points of zero angular velocity (i.e., blue regions) the angular velocity is roughly described by a sin function.
\label{fig:MeanAngSpeed_3000au}}
\end{figure}

We then calculate the on-sky angular speed ($\mu$) distribution as a function of sky position.
Figure~\ref{fig:MeanAngSpeed_3000au} shows the mean on-sky angular speed as a function of sky position
for objects with a semi-major axis of $a=3000$\,AU.
The solid line represents the ecliptic, and the black circle is the Sun position at the time of evaluation.
The two points of lowest mean angular speeds, $\mu$, can be seen on the ecliptic at an angular distance of 89\,deg from the Sun.
Using Equation~\ref{eqn:v_perp} with an Earth velocity of 29.8\,km\,s$^{-1}$, an elongation between 0 and 360\,deg, an Earth semi-major axis of 1\,AU, an object semi-major axis of 3000\,AU, and an object velocity of $V_{obj} = V_{\oplus}/\sqrt a$, yields a maximum angular speed of around 14\,$\mu$arcsec\,s$^{-1}$ for an object with zero eccentricity. The maximum angular speed, given an eccentricity of 0.6, goes up to around 35\,$\mu$arcsec\,s$^{-1}$.

\begin{figure}
\includegraphics[width=8cm]{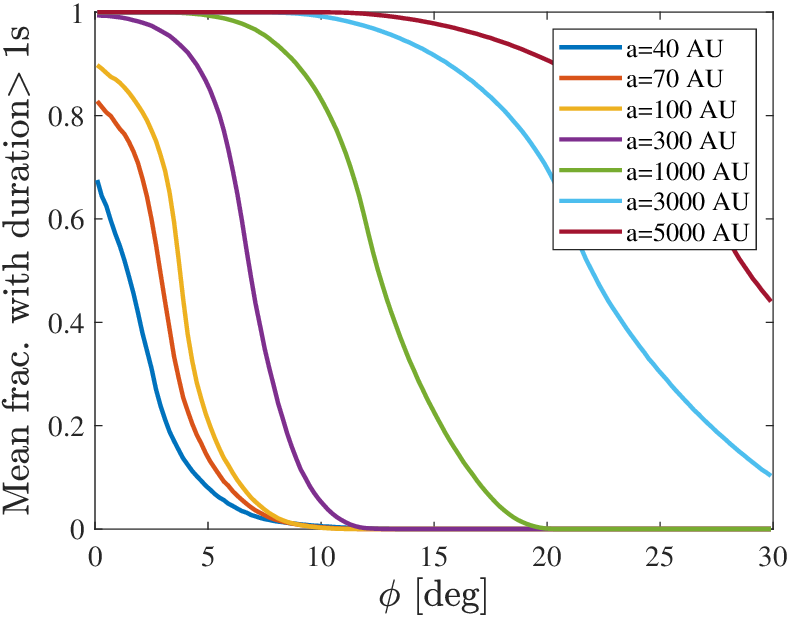}
\caption{Mean fraction of occultation events that have semi-duration larger than 1\,s,
as a function of angular distance from the point of zero angular speed ($\phi$).
This is calculated using the simulations
and orbital elements distribution discussed in the text. 
Different lines are for various semi-major axes (see legend).
\label{fig:MeanFraction_all_a_ExpTime1}}
\end{figure}

\begin{figure}
\includegraphics[width=8cm]{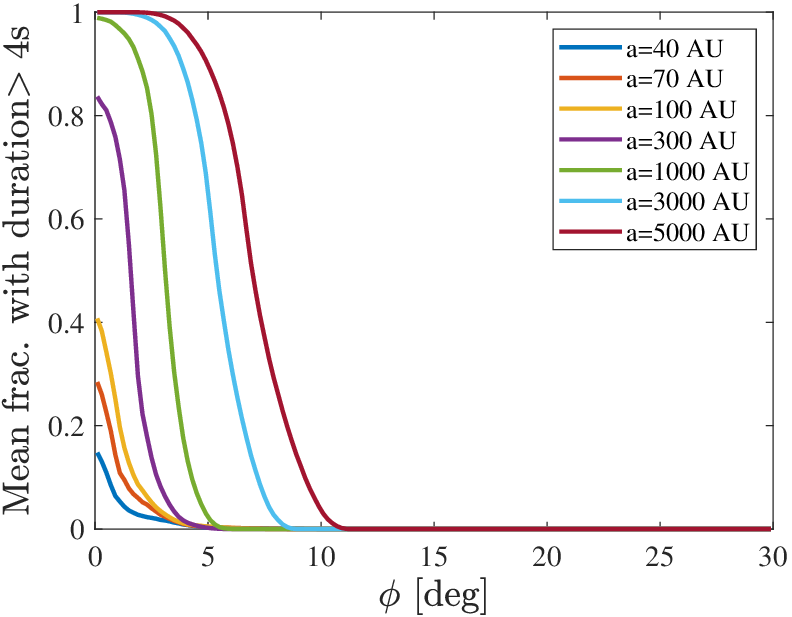}
\caption{Like Figure~\ref{fig:MeanFraction_all_a_ExpTime1}, but for an exposure time of 4\,s.
\label{fig:MeanFraction_all_a_ExpTime4}}
\end{figure}

Next, we calculate the fraction of objects, $f_{\rm occ}$ 
that have an occultation half-duration ($\theta_{\rm F}/\mu$) that is longer than some exposure time,
as a function of sky position. 
Based on these simulations, figures~\ref{fig:MeanFraction_all_a_ExpTime1}--\ref{fig:MeanFraction_all_a_ExpTime4} show the average fraction $f_{\rm occ}$ of occultations with semi-duration above 1\,s
and 4\,s, respectively, as a function of angular distance
from the point of zero angular speed ($\phi$), for several semi-major axes ($a$).



We next use Equation~\ref{eq:OccRate} to calculate the
occultation rate as a function of $a$, $q$, $\phi$, and the exposure time ($\Delta{t}$).
Since this parameter space is wide, we provide here several examples demonstrating the general trends.
In these calculations, we assume that to detect an event, we require that its Fresnel-radius crossing time will be longer than the exposure time (i.e., the events can be detected using two photometric data points).

\begin{figure}
\includegraphics[width=8cm]{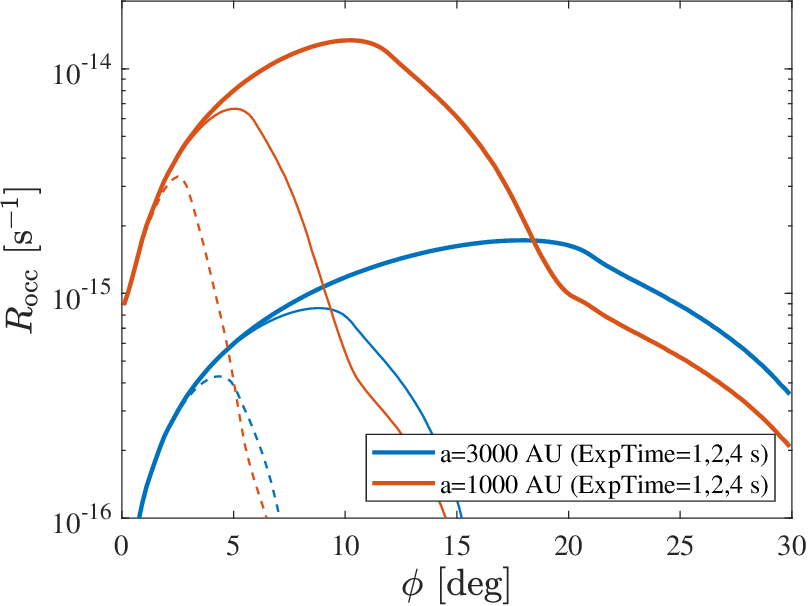}
\caption{The occultation rate (occultation probability per star per second) as a function of the angular distance from the point of minimum angular speed ($\phi$; note that this point depends on the semi-major axis), that can be detected using an exposure time of $\Delta{t}$\,seconds.
This is shown for semi-major axes of approximately 3000\,AU (blue lines) and 1000\,AU (orange lines).
For each semi-major axis, we show three lines corresponding to three exposure times of $\Delta{t}=1$\,s (solid heavy), 2\,s (solid thin), and 4\,s (dashed).
In all cases, we assumes $N_{>1\,{\rm km}}=10^{13}$ and $q=3$.
\label{fig:Rocc_phi_a_ExpTime}}
\end{figure}

In Figure~\ref{fig:Rocc_phi_a_ExpTime}, we present the rate of occultations whose semi-duration is longer than $\Delta{t}$\, seconds, as a function of the angular distance from the point of minimum angular speed ($\phi$; note that this point depends on the semi-major axis).
This is shown for semi-major axes of approximately 3000\,AU (blue lines) and 1000\,AU (orange lines).
For each semi-major axis, we show three lines corresponding to three exposure times of $\Delta{t}=1$\,s (solid heavy), 2\,s (solid thin), and 4\,s (dashed).

Optimization of a sky survey depends on several parameters, including the sky survey's available field of view, minimum exposure time, and limiting magnitude as a function of exposure time.
In addition, using the quadrature method, one can run out of sky quickly. For example, for KBO detection ($a\approx40$\,AU), the maximum angular distance from the point of minimum angular speed ($\phi$) in which one can detect occultations is only $\sim1$ (3)\,deg for exposure times of 4 (1)\,s (see Figure~\ref{fig:MeanFraction_all_a_ExpTime1}--\ref{fig:MeanFraction_all_a_ExpTime4}).

To demonstrate the order of magnitude capabilities of the quadrature method,
we use our simulations to estimate its detection capabilities for Tomo-e Gozen- (\citealt{Sako+2018SPIE_Tomo_e_Gozen_Overview})
and the Large Array Survey Telescope (LAST; \citealt{Ofek+2023PASP_LAST_Overview, BenAmi+2023PASP_LAST_Science}) like surveys.
Here we assume the surveys have a rectangular field centered on the point of minimum angular speed, and using the simulations, we calculate the event rate over the entire field of view.
In this case, the main parameter of the survey is the exposure time.

\begin{figure}
\includegraphics[width=8cm]{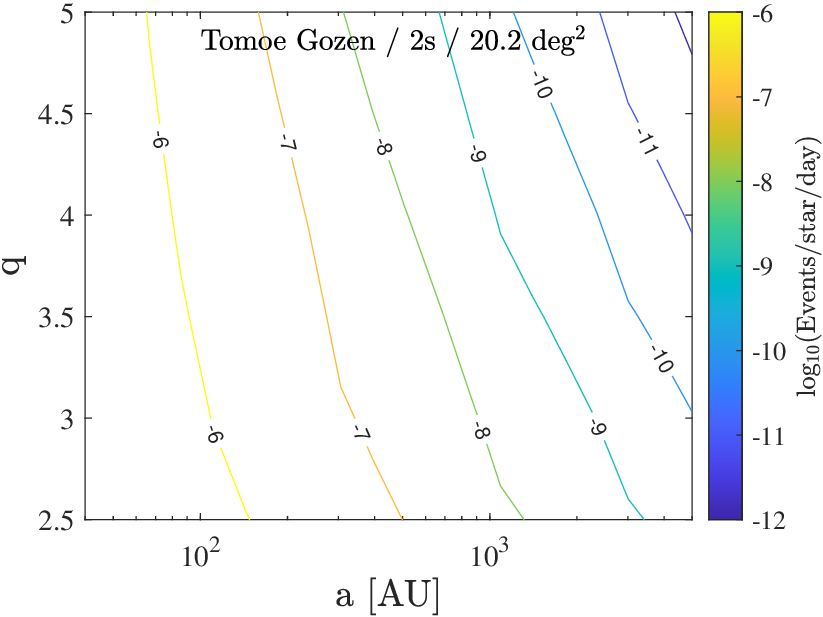}
\caption{The $\log_{10}$ of the expected mean occultations rate per star per 24\,hr of observing time, over a field of view of 20.25\,deg$^{2}$ centered on the point of minimum angular speed, and for Fresnel-radius crossing time of $>2$\,s.
These parameters are relevant for Tomo-e Gozen.
The rate is for an Oort-cloud-like population ($N_{>1}=10^{13}$).
We estimate that for KBOs the rates indicated in the Figure should be multiplied by about $10^{-2.5}$ (see text).
\label{fig:DetectionRatePerDay_a_q_TG_2s}}
\end{figure}

\begin{figure}
\includegraphics[width=8cm]{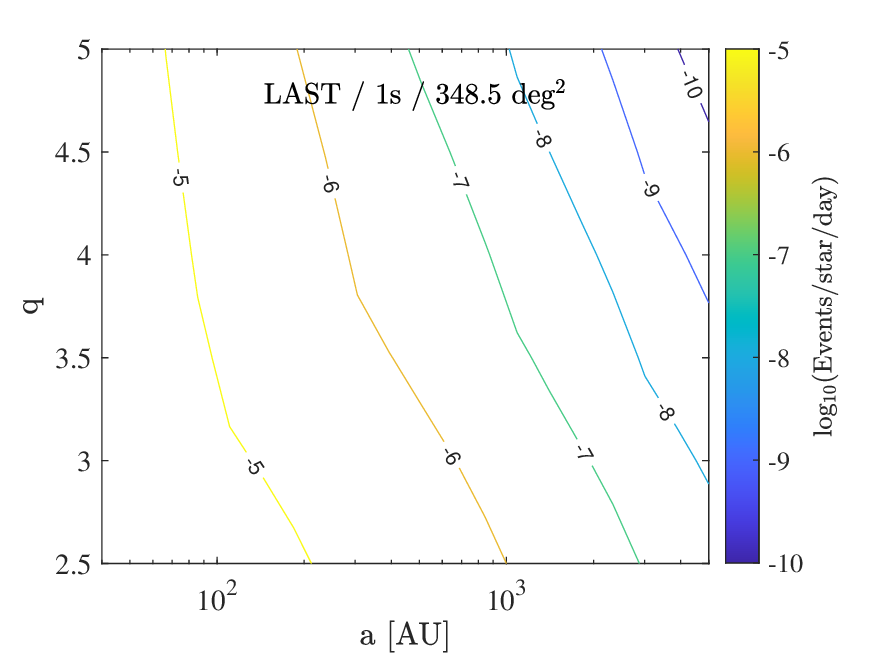}
\caption{Like Figure~\ref{fig:DetectionRatePerDay_a_q_TG_2s} but for a field of view of 348\,deg$^{2}$ centered on the point of minimum angular speed, and for a Fresnel-radius crossing time of $>1$\,s.
These parameters are relevant for LAST, with an exposure time of 1\,s.
\label{fig:DetectionRatePerDay_a_q_LAST_1s}}
\end{figure}

\begin{figure}
\includegraphics[width=8cm]{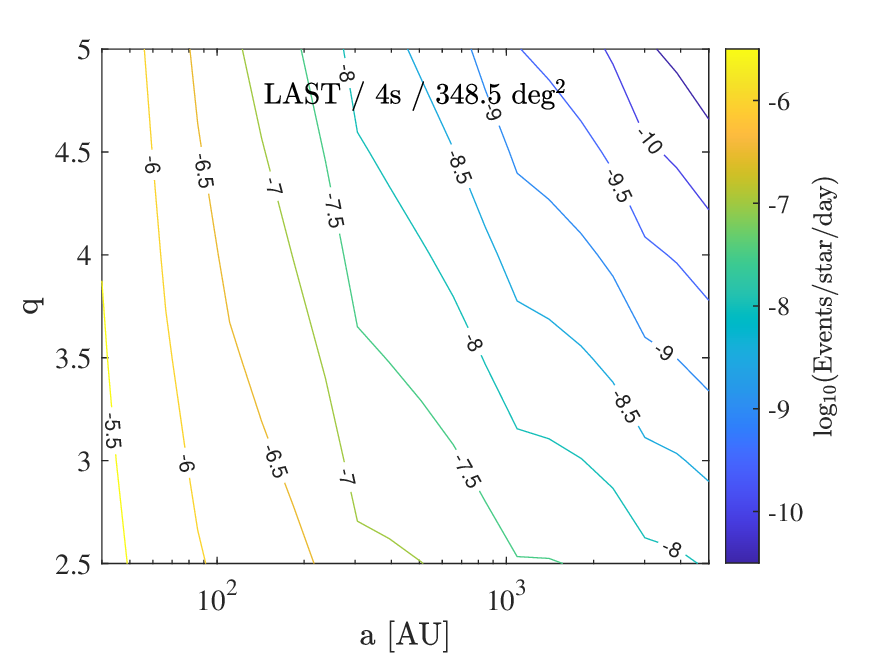}
\caption{Like Figure~\ref{fig:DetectionRatePerDay_a_q_TG_2s} but for a field of view of 348\,deg$^{2}$ centered on the point of minimum angular speed, and for a Fresnel-radius crossing time of $>4$\,s.
These parameters are relevant for LAST, with an exposure time of 4\,s.
\label{fig:DetectionRatePerDay_a_q_LAST_4s}}
\end{figure}

For Tomo-e Gozen, we assume a square field of view of $4.5\times4.5$\,deg$^{2}$ and exposure time of 2\,s. In this case, the occultation probability per day per suitable star in the field of view, as a function of $a$ and $q$ is shown in Figure~\ref{fig:DetectionRatePerDay_a_q_TG_2s}.
For LAST, we assume a field of view of 17.6 by 19.8 deg (48 telescopes).
Figure~\ref{fig:DetectionRatePerDay_a_q_LAST_1s} shows the occultation probability for LAST with 1\,s exposure time, while Figure~\ref{fig:DetectionRatePerDay_a_q_LAST_4s} is for 4\,s exposure time.

Next, to convert these plots to detection rates we need to know the number of observed stars within the field of view as a function of the exposure time.
For Tomo-e Gozen, we assume a limiting magnitude of about 18.5 in a 2\,s exposure, and we assume the number of stars is about $10^{4}$\,deg$^{-2}$ (90th upper percentile in Figure~\ref{fig:SorcesSmallerThanTF_vsMag}).
While for LAST, we assume a limiting magnitude of 17 in 1\,s, 17.7 in 2\,s, and 18.5 in 4\,s exposures.
The corresponding number of suitable stars are 3000,  8000, and 10,000\,deg$^{-2}$ (90th upper percentile in Figure~\ref{fig:SorcesSmallerThanTF_vsMag}), respectively.
From these numbers, as well as our previous assumptions, we can estimate the detection rate of Oort cloud objects using
Tomo-e Gozen and LAST.
We assume $a=10^{3}$\,AU and $q=3$ for these calculations.
For Tomo-e Gozen, the expected detection rate is $0.001$\,day$^{-1}$,
while for LAST, the rate is about $0.4$, $0.3$ and $0.06$\,day$^{-1}$,
for exposure times of 1, 2, and 4\,s, respectively.
From Figure~\ref{fig:DetectionRatePerDay_a_q_TG_2s}, and Equation~\ref{eq:OccRate}, we see that the rate of occultations roughly scales like
$\propto N_{\rm >1}$, $\propto a^{-2}$, and $\propto q^{-3.3}$.
These rates are estimates assuming $S/N=5$ for detection, and the sky background in the LAST site ($\approx20.8$\,mag\,arcsec$^{-2}$. However, for detection of an occultation, a source with somewhat larger $S/N$ may be required.
The number of suitable stars as a function of magnitude (Fig.~\ref{fig:SorcesSmallerThanTF_vsMag}), around magnitude 17.5, scales like $\propto 3\Delta{m}$, where $\Delta{m}$ is the change in the assumed limiting magnitude.
For example, if a $S/N=10$ is required, the rate will be decrease by a factor of about 2. However, if darker site may be used (e.g., $22$\,mag\,arcsec$^{-2}$), then the number of suitable sources may increase by a factor of about 2.

The parameters for these plots are relevant for an Oort cloud objects population.
To estimate the occultation rate for KBOs,
the rates in these plots should be multiplied by about $10^{-2.5}$.
The reasons for this reduction are:
(i) we expect about a $10^{-4}$ lower normalization of $N_{>1\,\rm km}$ for KBOs (\citealt{Schlichting+Ofek+2012_HST_KBO_Occultations}), compared with Oort cloud objects; 
(ii) The surface density of stars smaller than 42\,$\mu$arcsec (i.e., the Fresnel radius for the KBOs)
is about half an order of magnitude higher (Fig.~\ref{fig:SorcesSmallerThanTF_vsMag});
and (iii) since KBOs are concentrated near the ecliptic,
their surface density is about an order of magnitude higher
near the ecliptic compared with their all-sky mean surface density.
Finally, the larger Fresnel size of KBOs is already accounted for in Equation~\ref{eq:OccRate}.


Multiplying the rate, from Figure~\ref{fig:DetectionRatePerDay_a_q_TG_2s}--\ref{fig:DetectionRatePerDay_a_q_LAST_4s}, for KBOs (e.g., $a=40$\,AU; $q=3$)
by $10^{-2.5}$, we see that the expected detection rate for Oort cloud objects (with $a=10^{3}$\,AU; $q=3$) becomes
marginally higher than the detection rate for KBOs.
Therefore, unlike opposition observations which are expected to yield a higher occultation rate for KBOs, in comparison to Oort cloud objects,
the quadrature method may be slightly more efficient for detecting the Oort cloud.

\section{Conclusions}
\label{sec:Conclusion}

We analyze the likelihood of detecting Oort cloud objects using a few-second integration observations at quadrature.
This method was first suggested by \cite{Brown+Webster_1997MNRAS_KBO_QuadratureOccultationsDetection}
as a way to overcome the technical problem of short integration times
required for KBO detection by occultations (e.g., \citealt{Bailey1976_OccultationsKBO_idea}).
Here we show that the efficiency of this method for Oort cloud object detection
using small (sub-meter size) telescopes is high.
The reason for this is that while the longer occultation duration
near the quadrature points means lower occultation rates,
the longer integration times result in a large increase in the number
of stars that are smaller than the Fresnel size for Oort cloud objects.

This method potentially has another advantage -- the longer integration times
also mean higher resilience to intensity scintillation (e.g., \citealt{Young1967AJ_Photometry_Scintialtions_ReigerTheoryConfirmation}; \citealt{Osborn+2015MNRAS_Photometry_Scintilation}).
Nevertheless, using this method requires a new generation of survey telescopes
with a large field of view that are capable of observing at 
a few-second integration times with negligible dead time.
Examples of such surveys are the operational Tomo-e Gozen (\citealt{Sako+2018SPIE_Tomo_e_Gozen_Overview}),
the under-commissioning Large Array Survey Telescope (LAST;\citealt{Ofek+2023PASP_LAST_Overview, BenAmi+2023PASP_LAST_Science, Ofek+2023PASP_LAST_PipeplineI}),
and the planned Argus array (\citealt{Law+2022PASP_ArgusArray}).
A disadvantage of this method is that the sky area that can be used is $\sim1$\% of the celestial sphere.
A possible way to overcome this problem is to use multiple observatories 
that scan independent realizations of the sky. 
I.e., By changing the observatory's location, the Oort cloud objects occult at different sky locations.
The typical distance between such observatories
should be larger than the Fresnel size for the Oort cloud ($>10$\,km).

We thank an anonymous referee for useful comments.
E.O.O. is grateful for the support of
grants from the 
Willner Family Leadership Institute,
André Deloro Institute,
Paul and Tina Gardner,
The Norman E Alexander Family M Foundation ULTRASAT Data Center Fund,
Israel Science Foundation,
Israeli Ministry of Science,
Minerva,
NSF-BSF,
Israel Council for Higher Education (VATAT),
Sagol Weizmann-MIT,
Yeda-Sela,
and the Rosa and Emilio Segre Research Award.
This research was supported by the Institute for Environmental Sustainability (IES) and The André Deloro Institute for Space and Optics Research at the Weizmann Institute of Science.
S.A.S. is grateful for support by the Zuckerman Scholars Program.

\appendix
\section{Effect of extinction on angular size}
\label{Ap:ext}

\begin{figure}
\includegraphics[width=8cm]{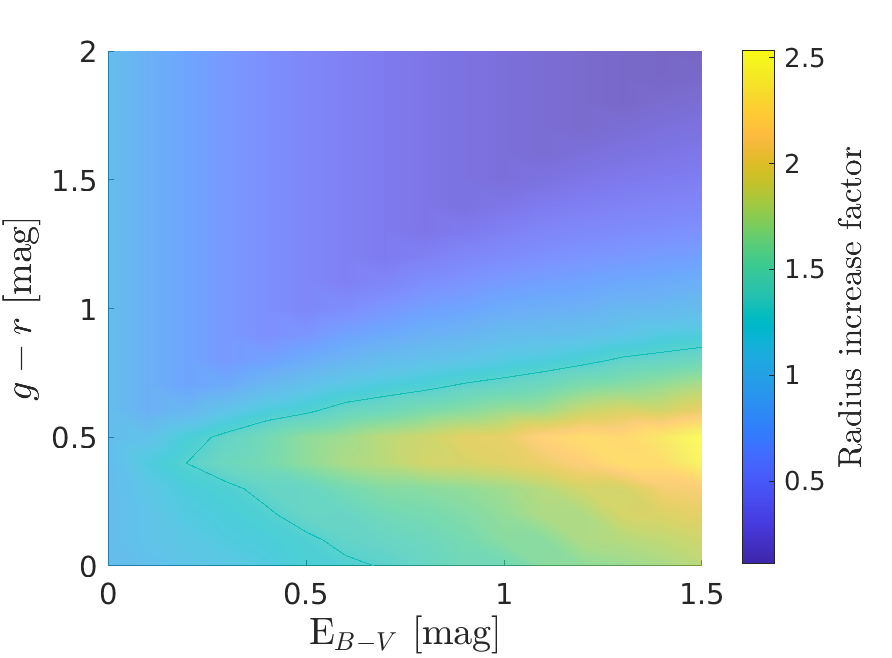}
\caption{The multiplication factor by which the angular size of a star is distorted if estimated based on the observed $g$ and $r$ filters,
as a function of the $E_{B-V}$ parameter, the $g-r$ color, and assuming $R_{V}=3.08$.
The solid green line shows the level at which the correction factor is 1.
\label{fig:E_color_size_R308}}
\end{figure}

Extinction and reddening complicate the estimation of stars' angular sizes.
Specifically, reddening changes the temperature estimation,
and extinction affects the star's flux ($F$).
Since $\theta_{*}\propto T^{-2} F^{1/2}$, both are important.
Figure~\ref{fig:E_color_size_R308} shows
the multiplication factor by which the fitted angular size of a star should be multiplied in order to get its real angular size, as a function of the $E_{B-V}$ parameter
and the star $g-r$ color.
Here we assume the measured angular radius of the stars is based on fitting the $g$ and $r$ magnitudes
with a black-body curve.
We also assume $R_{V}=3.08$
(\citealt{Cardelli+1989_Extinction}).
This plot demonstrates that for a wide range of the parameter space, 
the angular size correction factor is below 1 
(i.e., the actual angular size is smaller than estimated by the fit).
Therefore, we conclude 
that the estimate for the number of stars with an angular size smaller than 5\,$\mu$arcsec is reasonable.

\bibliography{papers.bib}

\begin{thebibliography}{}
\expandafter\ifx\csname natexlab\endcsname\relax\def\natexlab#1{#1}\fi

\bibitem[{{Arimatsu} {et~al.}(2019){Arimatsu}, {Tsumura}, {Usui}, {Shinnaka},
  {Ichikawa}, {Ootsubo}, {Kotani}, {Wada}, {Nagase}, \&
  {Watanabe}}]{Arimatsu+2019_AkmSizedKBO_Occultation_GroundBasedTelescope}
{Arimatsu}, K., {Tsumura}, K., {Usui}, F., {et~al.} 2019, Nature Astronomy, 3,
  301

\bibitem[{{Bailey}(1976)}]{Bailey1976_OccultationsKBO_idea}
{Bailey}, M.~E. 1976, \nat, 259, 290

\bibitem[{{Ben-Ami} {et~al.}(2023){Ben-Ami}, {Ofek}, {Polishook},
  {Franckowiak}, {Hallakoun}, {et~al.}}]{BenAmi+2023PASP_LAST_Science}
{Ben-Ami}, S., {Ofek}, E.~O., {Polishook}, D., {et~al.} 2023, arXiv e-prints,
  arXiv:2304.02719

\bibitem[{{Bianco} {et~al.}(2009){Bianco}, {Protopapas}, {McLeod}, {Alcock},
  {Holman}, \& {Lehner}}]{Bianco+2009_OccultationsKBO_MMT_search}
{Bianco}, F.~B., {Protopapas}, P., {McLeod}, B.~A., {et~al.} 2009, \aj, 138,
  568

\bibitem[{{Brasser} {et~al.}(2006){Brasser}, {Duncan}, \&
  {Levison}}]{Brasser+2006_OortCloudFormation_EmbededStarClustersEnvironment}
{Brasser}, R., {Duncan}, M.~J., \& {Levison}, H.~F. 2006, Icarus, 184, 59

\bibitem[{{Brown} \&
  {Webster}(1997)}]{Brown+Webster_1997MNRAS_KBO_QuadratureOccultationsDetection}
{Brown}, M.~J.~I., \& {Webster}, R.~L. 1997, \mnras, 289, 783

\bibitem[{{Cardelli} {et~al.}(1989){Cardelli}, {Clayton}, \&
  {Mathis}}]{Cardelli+1989_Extinction}
{Cardelli}, J.~A., {Clayton}, G.~C., \& {Mathis}, J.~S. 1989, \apj, 345, 245

\bibitem[{{Chambers} {et~al.}(2016){Chambers}, {Magnier}, {Metcalfe},
  {Flewelling}, {Huber}, {Waters}, {Denneau}, {Draper}, {Farrow}, {Finkbeiner},
  {Holmberg}, {Koppenhoefer}, {Price}, \& {Rest}}]{Chambers+2016_PS1_Surveys}
{Chambers}, K.~C., {Magnier}, E.~A., {Metcalfe}, N., {et~al.} 2016, arXiv
  e-prints, arXiv:1612.05560

\bibitem[{{Dones} {et~al.}(2004){Dones}, {Weissman}, {Levison}, \&
  {Duncan}}]{Dones+2004_OortCloudFormation}
{Dones}, L., {Weissman}, P.~R., {Levison}, H.~F., \& {Duncan}, M.~J. 2004, in
  Astronomical Society of the Pacific Conference Series, Vol. 323, Star
  Formation in the Interstellar Medium: In Honor of David Hollenbach, ed.
  D.~{Johnstone}, F.~C. {Adams}, D.~N.~C. {Lin}, D.~A. {Neufeeld}, \& E.~C.
  {Ostriker}, 371

\bibitem[{{Duncan} {et~al.}(1987){Duncan}, {Quinn}, \&
  {Tremaine}}]{Duncan+1987_OortCloudFormation}
{Duncan}, M., {Quinn}, T., \& {Tremaine}, S. 1987, \aj, 94, 1330

\bibitem[{{Gaudi}(2004)}]{Gaudi2004ApJ_Kepler_KuiperBelt_Occultations}
{Gaudi}, B.~S. 2004, \apj, 610, 1199

\bibitem[{{Harding} {et~al.}(2016){Harding}, {Hallinan}, {Milburn}, {Gardner},
  {Konidaris}, {Singh}, {Shao}, {Sandhu}, {Kyne}, \&
  {Schlichting}}]{Harding+2016_CHIMERA_P200}
{Harding}, L.~K., {Hallinan}, G., {Milburn}, J., {et~al.} 2016, \mnras, 457,
  3036

\bibitem[{{Huang} {et~al.}(2021){Huang}, {Lehner}, {Granados Contreras},
  {Castro-Chac{\'o}n}, {Chen}, {Alcock}, {Alvarez-Santana}, {Cook}, {Geary},
  {Guerrero Pe{\~n}a}, {Hern{\'a}ndez-{\'A}guila}, {Hern{\'a}ndez-Valencia},
  {Karr}, {Kavelaars}, {Norton}, {Nu{\~n}ez}, {Ochoa}, {Reyes-Ruiz},
  {S{\'a}nchez}, {Silva}, {Szentgyorgyi}, {Wang}, {Yen}, \&
  {Zhang}}]{Huang+2021PASP_TAOS_II_Occultations_KBOs_Overview}
{Huang}, C.-K., {Lehner}, M.~J., {Granados Contreras}, A.~P., {et~al.} 2021,
  \pasp, 133, 034503

\bibitem[{{Law} {et~al.}(2022){Law}, {Corbett}, {Galliher}, {Gonzalez},
  {Vasquez}, {Walters}, {Machia}, {Ratzloff}, {Ackley}, {Bizon}, {Clemens},
  {Cox}, {Eikenberry}, {Howard}, {Glazier}, {Mann}, {Quimby}, {Reichart}, \&
  {Trilling}}]{Law+2022PASP_ArgusArray}
{Law}, N.~M., {Corbett}, H., {Galliher}, N.~W., {et~al.} 2022, \pasp, 134,
  035003

\bibitem[{{Nir} {et~al.}(2023{\natexlab{a}}){Nir}, {Ofek}, {Polishook},
  {Zackay}, \& {Ben-Ami}}]{Nir+Ofek+2023_WFAST_KBO_Search2020_2021}
{Nir}, G., {Ofek}, E.~O., {Polishook}, D., {Zackay}, B., \& {Ben-Ami}, S.
  2023{\natexlab{a}}, \mnras, 526, 43

\bibitem[{{Nir} {et~al.}(2023{\natexlab{b}}){Nir}, {Ofek}, \&
  {Zackay}}]{Nir+Ofek+2023_TNO_Occultations_Pipeline}
{Nir}, G., {Ofek}, E.~O., \& {Zackay}, B. 2023{\natexlab{b}}, RAS Techniques
  and Instruments, 2, 567

\bibitem[{{Nir} {et~al.}(2021){Nir}, {Ofek}, {Ben-Ami}, {Segev}, {Polishook},
  {Hershko}, {Diner}, {Manulis}, {Zackay}, {Gal-Yam}, \&
  {Yaron}}]{Nir+Ofek+2021_WFAST}
{Nir}, G., {Ofek}, E.~O., {Ben-Ami}, S., {et~al.} 2021, \pasp, 133, 075002

\bibitem[{{Ofek}(2014)}]{Ofek2014_MAAT}
{Ofek}, E.~O. 2014, {MATLAB package for astronomy and astrophysics},
  ascl:1407.005

\bibitem[{{Ofek} \&
  {Ben-Ami}(2020)}]{Ofek+BenAmi2020_Grasp_SkySurvrys_CostEffectivness}
{Ofek}, E.~O., \& {Ben-Ami}, S. 2020, arXiv e-prints, arXiv:2011.04674

\bibitem[{{Ofek} {et~al.}(2023{\natexlab{a}}){Ofek}, {Ben-Ami}, {Polishook},
  {Segre}, {Blumenzweig}, {et~al.}}]{Ofek+2023PASP_LAST_Overview}
{Ofek}, E.~O., {Ben-Ami}, S., {Polishook}, D., {et~al.} 2023{\natexlab{a}},
  \pasp, 135, 065001

\bibitem[{{Ofek} \& {Nakar}(2010)}]{Ofek+2010_OortOccultation_Kepler}
{Ofek}, E.~O., \& {Nakar}, E. 2010, \apj, 711, L7

\bibitem[{{Ofek} {et~al.}(2023{\natexlab{b}}){Ofek}, {Shvartzvald}, {Sharon},
  {Tishler}, {Elhanati}, {et~al.}}]{Ofek+2023PASP_LAST_PipeplineI}
{Ofek}, E.~O., {Shvartzvald}, Y., {Sharon}, A., {et~al.} 2023{\natexlab{b}},
  \pasp, 135, 124502

\bibitem[{{Oort}(1950)}]{Oort1950_TheOortCloud}
{Oort}, J.~H. 1950, \bain, 11, 91

\bibitem[{{Osborn} {et~al.}(2015){Osborn}, {F{\"o}hring}, {Dhillon}, \&
  {Wilson}}]{Osborn+2015MNRAS_Photometry_Scintilation}
{Osborn}, J., {F{\"o}hring}, D., {Dhillon}, V.~S., \& {Wilson}, R.~W. 2015,
  \mnras, 452, 1707

\bibitem[{{Roques} {et~al.}(2006){Roques}, {Doressoundiram}, {Dhillon},
  {Marsh}, {Bickerton}, {Kavelaars}, {Moncuquet}, {Auvergne}, {Belskaya},
  {Chevreton}, {Colas}, {Fernandez}, {Fitzsimmons}, {Lecacheux}, {Mousis},
  {Pau}, {Peixinho}, \& {Tozzi}}]{Roques+2006_KBO_Occultations_Search}
{Roques}, F., {Doressoundiram}, A., {Dhillon}, V., {et~al.} 2006, \aj, 132, 819

\bibitem[{{Sako} {et~al.}(2018){Sako}, {Ohsawa}, {Takahashi}, {Kojima}, {Doi},
  {Kobayashi}, {Aoki}, {Arima}, {Arimatsu}, {Ichiki}, {Ikeda}, {Inooka}, {Ita},
  {Kasuga}, {Kokubo}, {Konishi}, {Maehara}, {Matsunaga}, {Mitsuda}, {Miyata},
  {Mori}, {Morii}, {Morokuma}, {Motohara}, {Nakada}, {Okumura}, {Sarugaku},
  {Sato}, {Shigeyama}, {Soyano}, {Tanaka}, {Tarusawa}, {Tominaga}, {Totani},
  {Urakawa}, {Usui}, {Watanabe}, {Yamashita}, \&
  {Yoshikawa}}]{Sako+2018SPIE_Tomo_e_Gozen_Overview}
{Sako}, S., {Ohsawa}, R., {Takahashi}, H., {et~al.} 2018, in Society of
  Photo-Optical Instrumentation Engineers (SPIE) Conference Series, Vol. 10702,
  Ground-based and Airborne Instrumentation for Astronomy VII, ed. C.~J.
  {Evans}, L.~{Simard}, \& H.~{Takami}, 107020J

\bibitem[{{Schlafly} {et~al.}(2012){Schlafly}, {Finkbeiner}, \&
  {Juric}}]{Schlafly+2012_PS1_PhotometricCalibration}
{Schlafly}, E., {Finkbeiner}, D.~P., \& {Juric}, M. 2012, in American
  Astronomical Society Meeting Abstracts, Vol. 219, American Astronomical
  Society Meeting Abstracts \#219, 428.16

\bibitem[{{Schlichting} {et~al.}(2009){Schlichting}, {Ofek}, {Wenz}, {Sari},
  {Gal-Yam}, {Livio}, {Nelan}, \&
  {Zucker}}]{Schlichting+Ofek+2009_HST_KBO_Occultations}
{Schlichting}, H.~E., {Ofek}, E.~O., {Wenz}, M., {et~al.} 2009, Nature, 462,
  895

\bibitem[{{Schlichting} {et~al.}(2012){Schlichting}, {Ofek}, {Sari}, {Nelan},
  {Gal-Yam}, {Wenz}, {Muirhead}, {Javanfar}, \&
  {Livio}}]{Schlichting+Ofek+2012_HST_KBO_Occultations}
{Schlichting}, H.~E., {Ofek}, E.~O., {Sari}, R., {et~al.} 2012, \apj, 761, 150

\bibitem[{{Soumagnac} \& {Ofek}(2018)}]{Soumagnac+Ofek2018_catsHTM}
{Soumagnac}, M.~T., \& {Ofek}, E.~O. 2018, \pasp, 130, 075002

\bibitem[{{Young}(1967)}]{Young1967AJ_Photometry_Scintialtions_ReigerTheoryConfirmation}
{Young}, A.~T. 1967, \aj, 72, 747

\bibitem[{{Zhang} {et~al.}(2013){Zhang}, {Lehner}, {Wang}, {Wen}, {Wang},
  {King}, {Granados}, {Alcock}, {Axelrod}, {Bianco}, {Byun}, {Chen}, {Coehlo},
  {Cook}, {de Pater}, {Kim}, {Lee}, {Lissauer}, {Marshall}, {Protopapas},
  {Rice}, \& {Schwamb}}]{Zhang+2013_TAOS_I_Results7years}
{Zhang}, Z.~W., {Lehner}, M.~J., {Wang}, J.~H., {et~al.} 2013, \aj, 146, 14

\end{thebibliography}
\bibliographystyle{apj}

\end{document}